# Room Temperature Quantum Control of N-Donor Electrons at Si/SiO$_2$ Interface

Soumya Chakraborty[1], Arup Samanta*[1,2]

[1]Quantum/Nano-science and technology Lab, Department of Physics, Indian Institute of Technology Roorkee, Roorkee-247667, Uttarakhand, India
[2]Centre for Nanotechnology, Indian Institute of Technology Roorkee, Roorkee-247667, Uttarakhand, India
*arup.samanta@ph.iitr.ac.in

**Abstract:**

The manuscript theoretically discusses various important aspects for donor atom based single qubit operations in silicon (Si) quantum computer architecture at room temperature using a single nitrogen (N) deep-donor close to the Si/SiO$_2$ interface. Quantitative investigation of room temperature single electron shuttling between a single N-donor atom and the interface is the focus of attention under the influence of externally applied electric and magnetic field. To apprehend the realistic experimental configurations, central cell correction along with effective mass approach is adopted throughout the study. Furthermore, a detailed discussion currently explores the significant time scales implicated in the process and their suitability for experimental purposes. Theoretical estimates are also provided for all the external fields required to successfully achieve coherent single electron shuttling and their stable maintenance at the interface as required. The results presented in this work offer practical guidance for quantum electron control using N-donor atoms in Si at room temperature.

## 1. Introduction

The revolutionary emergence of quantum computing has the potential to transform information processing to an enormous extent. Dopant atom based silicon (Si) devices have been extensively studied as a practical platform for quantum computing.[1-4] Silicon offers numerous advantages, including well-established fabrication techniques,[5] long coherence times,[6] and the potential for integration with existing complementary metal-oxide-semiconductor (CMOS) technology. These factors establish silicon-based devices as a promising avenue for developing spin or charge qubit architectures which, in a sense, are the backbone of quantum computers. The initial proposal for doped Si based quantum computer architecture was based on donor nuclear spins.[2] Further, single electronic spin as qubit is also being extensively studied.[7-9] Subsequently, donor based charge qubit has also been proposed.[10] Both of the nuclear and electronic spins in Si have large coherence time which is essential for practical quantum computing. The coherence times (T$_2$) in natural silicon for phosphorous-donor's nuclear spins[11-12] and electronic spins[11,13,14] are in the order of few mili-second (ms) and micro-second (μs) respectively. Whereas, the same for electronic charge is a few μs.[15,16] These time scales can further be enhanced by a few fold with isotope purification.[17] Although having shorter T$_2$, charge qubits are far easier to manipulate externally. Moreover, single electronic charge can easily be detected with state-of-the-art single-electron tunneling (SET) devices[18-23] as compared to spin qubit detection.[24] In addition, charge-based quantum operations can be achieved using conventional gate designs[25] and standard CMOS processes, facilitating their integration into existing semiconductor technology which make them a suitable candidate for qubit operation. Interfacial double-donor system[26] or the donor-interface system plays an intriguing role towards single electron shuttling for charge qubit operations.[27-29] Meticulous theoretical studies regarding this context with phosphorous (P) single-donor[27,28] and different double-donors (e.g., Te, Se, and Bi)[29] have been performed in the last two decades. The donor within silicon crystal acts as a localized charge source whereas the interface, typically composed of a silicon-silicon dioxide (Si/SiO$_2$) structure in MOSFET or Fin-FET configuration, serves as a triangular potential well that allows confinement and manipulation of single electrons between the donor and the interface well. This donor-interface system facilitates controlled transfer and storage of quantum information encoded in charge qubits.[30-32] So far, such sophisticated electron shuttling had been limited only up to sub-Kelvin temperatures because achieving high-temperature single electron shuttling, pillars of room temperature quantum computers, presents significant challenges. Elevated temperatures introduce the additional effect of thermal energy, leading to higher thermal fluctuations and reduced coherence time of qubits. Maintaining stability and coherence of charge qubits at higher temperatures are crucial for practical quantum computing applications. Several studies have highlighted the difficulties associated with such high-temperature single electron shuttling. For instance, the deteriorating effect of increased thermal noise on the effective single electron tunneling rate and overall device performance, in silicon-based systems, at elevated temperature.[33-35] To address the challenges associated with high-temperature single electron shuttling, recent studies have explored various strategies. These include dielectric and quantum confinement,[36] multi-dopant cluster,[37] different double donors,[29,38] and donor-vacancy interaction to mitigate the impact of thermal noise.[39,40] But, all these enormous efforts have either been practically realized only up to few tens of Kelvin temperatures or have technologically non-viable for large scale fabrication.

Here, we propose and theoretically investigate Nitrogen (N) deep-donor based device that will allow single electron shuttling between N-donor and Si/SiO$_2$ interface at room temperature. The valley-orbit interaction in Si due to the six valley degeneracy deepens the ground state (A$_1$) energy of N-donor almost around 190 meV below the conduction band.[41] For comparison, the same for P-donor in Si is only ~45 meV below the conduction band in bulk systems. The T$_2$ and E states, for both of the P and N-donor, remains around thirty meV below the conduction band being donor independent.[42] The A$_1$ state energy can further be enhanced in nano-dimensional devices due to various confinement effects,[43-44] which is reported in our recent paper along with the suitability of N-donor for single electron transistor at room temperature.[41] This huge energy difference between the ground and excited states in N-donor will reduce the thermal fluctuation greatly leading towards increased thermal stability. Moreover, it was experimentally demonstrated that the coherence time can be significantly enhanced by deepening the donor potential,[38,45,46] which can in turn be obtained through exploitation of multi donor cluster formation or by using deep-donor. So, it is expected that the deep nature of N-donor will enhance the coherence time. Another important aspect for such donor-based system in Si is the presence of the inherent interface states and their interference on device operation. Such interface states, for Si/SiO$_2$ interface, have density maximum ($\approx 4 \times 10^{10}$ cm$^{-2}$eV$^{-1}$) at energy values around 0.25 eV and 0.85 eV from the Si valence band[47] or about 0.2 eV away at both sides from Si midgap.[48] They also possess significant density around 10 meV to 40 meV below the Si conduction band minimum.[47,48] Energetically, the maximum of these interface states are almost 100 meV ($\approx 4$ k$_B$T at T $\approx 300$ K) away from the N-donor ground state level. Such large energy separation of the N-donor ground state and the inherent interface states will restrain any significant interaction between them. So, these states should not have any pernicious effect on the device performance. In addition, if we placed single N-dopant within a dimension 10 nm × 10 nm × 20-30 nm and applied proper passivation technique during fabrication, it can limit the inherent Si surface state density, as low as, up to the order of $10^{10}$ cm$^{-2}$eV$^{-1}$ and better.[49-51] In such situation, probability of such surface states per devices is only 1%. The interaction of interface state with the donor ground state can be avoided for most of the devices. However, if there is any interaction between the inherent interface state and the donor, we can see additional trapping and de-trapping phenomena. Such phenomena may occur in rare cases. Hence, N-donor could be a perfect candidate for room temperature single electron shuttling.

Here, we analyze the possible room temperature manipulation of donor electrons, using N-deep-donor, between the donor and interface by means of externally applied uniform electric and magnetic fields. In section 2, we discuss in detail the donor and interface ground states and their behavior under externally applied fields within the isolated donor model along with a brief formalism of electron shuttling. In section 3, we extensively analyze the electron shuttling between the donor and interface ground states at room temperature. Section 4 summarizes and concludes our study.

## 2. Donor Interface System

### 2.1. Formalism

To study and analyze the electron shuttling between donor and Si/SiO$_2$ interface at room temperature, we first consider a device as shown schematically in Figure 1. The local surface gate (G) is positioned exactly above the single deep N-donor to manipulate the electron between the donor potential and the externally created interface well by applied electric Field. The electric (**F**) and magnetic (**B**) fields are essential for the operation of these devices. Electric field can be applied through the top gate G. The magnetic field parallel to **F** can also be applied directly to the devices. The conduction band of Si inherits six equivalent minima along the $\Delta$ lines at about 85% to the Brillouine zone boundary from the center. To rationalize the problem, we adopt single valley effective mass formalism.[52-55] Further, we have incorporated a central cell correction potential of the form $Ce^{-fr} - gre^{-fr}$ to the donor potential that will make the Hamiltonian of the composite system donor species dependent. Here, '$f$' describes the range of the potential, and '$1/f$' is twenty times the difference between the atomic radius of donor and host material. '$C$' and '$g$' determine the strength of the species dependent potential. With these considerations, the Hamiltonian for the composite system can be described as

$$H = T + eFz - \frac{e^2}{r\varepsilon_{Si}}\left(1 - Ce^{-f\sqrt{\rho^2+z^2}} + gre^{-f\sqrt{\rho^2+z^2}}\right) + \frac{e^2 Q}{\varepsilon_{Si}\sqrt{\rho^2+(z+2d)^2}} - \frac{e^2 Q}{4\varepsilon_{Si}(z+d)} \quad (1)$$

The effect of the applied magnetic field, in terms of magnetic vector potential, $\mathbf{A} = \frac{B(y,-x,0)}{2}$, is included in the kinetic term (T) as

$$T = \sum_{\zeta=x,y,z} \hbar^2/(2m_\zeta)\,[i\partial/\partial\zeta + eA_\zeta/(\hbar c)]^2$$

The effective masses in Si are, $m_x = m_y = m_\perp = 0.191\, m_e$; $m_z = m_\parallel = 0.916\, m_e$. The second term in Equation (1) is the electric field linear potential, whereas the third one is the donor Coulomb potential. The last two terms are the donor and electron image terms respectively. $Q = (\varepsilon_{SiO_2} - \varepsilon_{Si})/(\varepsilon_{SiO_2} + \varepsilon_{Si})$, where $\varepsilon_{SiO_2}= 3.8$ and $\varepsilon_{Si} = 11.4$. To facilitate Equation (1), we rewrite it as:

$$H = -\frac{\partial^2}{\partial x^2} - \frac{\partial^2}{\partial y^2} - \gamma \frac{\partial^2}{\partial z^2} + \frac{1}{4}\mu^2\rho^2 + i\mu(y\partial_x - x\partial_y) + keFz - \frac{2}{r}(1 - Ce^{-fr} + gre^{-fr}) + \frac{2Q}{\sqrt{\rho^2+(z+2d)^2}} - \frac{Q}{2(z+d)}, \quad (2)$$

in rescaled atomic units. $a^* = \hbar^2 \varepsilon_{Si}/m_\perp e^2 = 3.156$ nm; $R_y^* = m_\perp e^4/2\hbar^2\varepsilon_{Si}^2 = 19.98$ meV; $\gamma = m_\perp/m_\parallel$; $\mu^2 = a^{*4}/\lambda_B^4$; $\lambda_B = \sqrt{\hbar/eB}$; $k = 3.89 \times 10^{-7}\varepsilon_{Si}^3(m_e/m_\perp)^2$cm/kV.

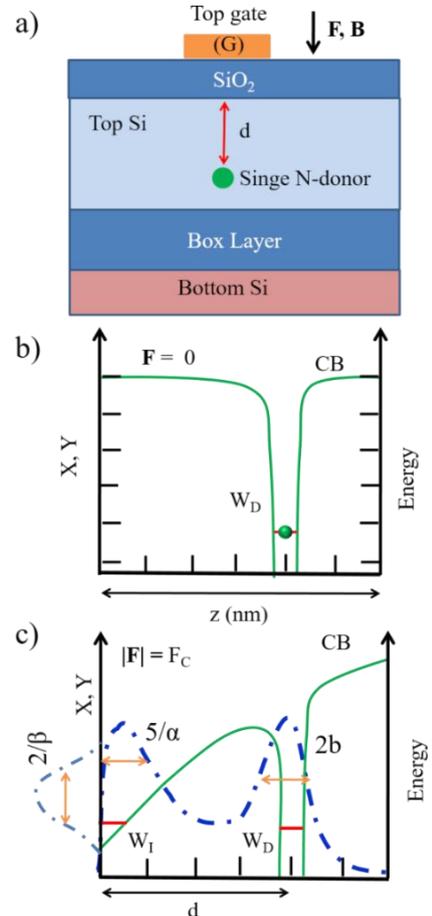

**Figure 1.** a) Schematic view of the system under study. A Nitrogen donor is placed at a depth $d$ from the Si/SiO$_2$ interface (001) exactly under the top control gate. **F**, **B** are applied perpendicular to the top surface, along Z axis. b) Schematic potential profile of the system without any external fields. c) External electric field dependent double well

configuration. The blue dashed line represents schematic electronic wave function distribution at $F = F_C$ and the corresponding ground state energies at both well.

We consider a single N-donor within Si matrix at a depth $d$ away from an impenetrable Si/SiO$_2$ interface. In absence of any external fields, the electron is bound to the donor potential well (W$_D$) ground state as shown in Figure 1b and is characterized by wave function $\Psi_D$. Upon turning on an external electric field $F$ along the z direction, as shown in Figure 1c, a triangular potential well (W$_I$) is formed at the interface. The corresponding ground state is characterized by the wave function $\Psi_I$. The Hamiltonian in Equation (2) is then written as,

$$H = \begin{bmatrix} H_{DD} & H_{DI} \\ H_{ID} & H_{II} \end{bmatrix},$$

in the non-orthogonal basis set $\{\Psi_D, \Psi_I\}$. Direct diagonalization of $H$ provides two eigen-energies $E^+$ and $E^-$ with eigenstate $\Psi^+$ and $\Psi^-$, respectively that show anti-crossing behavior with a minimum energy gap when $H_{DD} = H_{II}$. This point gives the characteristic electric field $F_C$ at which electron will shuttle from donor to the interface. Proper adjustment of $F$ and an externally applied magnetic field $B$ parallel to $F$ can shuttle the electron between N-donor and interface well reversibly, discussed in detail in Sec. 3.

## 2.2. Donor Ground State $\Psi_D$

To calculate the isolated donor ground state energy, we variationally solve the Hamiltonian,

$$H_D = T + V_D \quad (3)$$

With $V_D = -\frac{2}{r}(1 - Ce^{-fr} + gre^{-fr})$, in the basis $\Psi_D$. We chose anisotropic waveform,[27,56]

$$\Psi_D = N(z+d)e^{-\sqrt{\rho^2/a^2 + z^2/b^2}} \quad (4)$$

Where, $N = 1/\sqrt{\pi a^2 b (d^2 + b^2 - \frac{b}{2}(\frac{d}{2} + b)e^{-2d/b})}$.

$a$ and $b$ are the variational parameters to minimize the ground state energy towards the literature specified values.

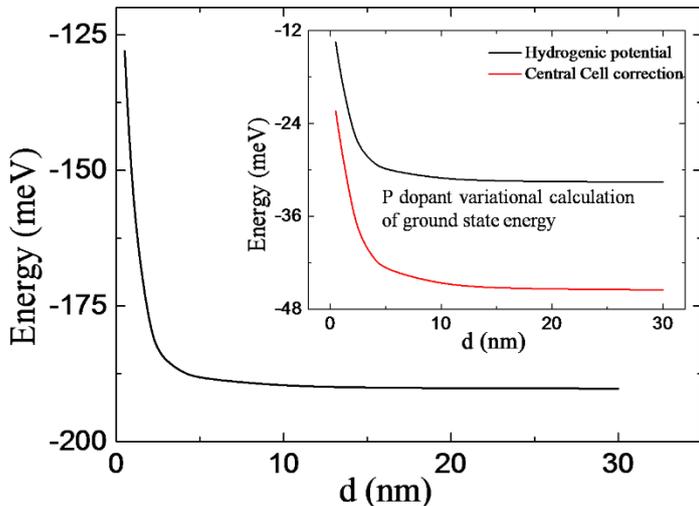

**Figure 2.** Variation of single N-donor ground state energy $\langle \psi_D|H_D|\psi_D\rangle$, calculated variationally using the wave function $\Psi_D$ and central cell corrected Hamiltonian $H_D$, with the distance from interface, $d$. The inset shows the ground state energy vs. $d$ for an isolated P atom placed within the system. The red solid line exhibits for central cell corrected Hamiltonian for isolated P atom in bulk Si to match experimental donor ground state energy of 45 meV; whereas, the black line in the inset defines a Hamiltonian without central cell correction term.

In Figure 2, we show the variational results for nitrogen ground state energy calculated at different depths from the interface. The inset depicts the same results for a shallow Phosphorous (P) donor for comparison. The values of the parameters '$C$' and '$g$' are chosen in such a way that matches the variational values of $a$ and $b$ as close as possible to the atomic Bohr radius of the corresponding species. Table 1 lists the values of different parameters used to minimize the ground state energy of a shallow P-donor and deep N-donor within the system under study, individually. Strong dependence of energy on $d$ is observed for smaller depths of the donor atom from the interface contrary to the case of medium to larger depths where the ground state energy remains almost constant towards the bulk value.

Moreover, the external electric and magnetic fields do not affect the donor levels significantly. For $d = 11$ nm and for $F = F_C$, the electric field induced potential energy is given as $<\psi_D|keFz|\psi_D> = 0.0859R_y^*$, compared to the bulk ground state energy of $9.5R_y^*$ for N-donor in Si. Thus the shift in donor ground state energy with $F$ is negligible. Similarly, to achieve a magnetic length of the order of Bohr radius $a = 1.05$ nm, $B \sim 600$ T is required. The shift in energy with $B$ can be quantified as[27,57]

$$\Delta E_B = \frac{\bar{r}^4}{4\lambda_B^4}R_y^* \quad (5)$$

So, we can clearly say that the donor ground state energy remains unaffected by the range of applied $E$ and $B$ that will be considered within this study.

| Donor | a (nm) | B (nm) | C | G | F (nm$^{-1}$) | Energy (meV) |
|---|---|---|---|---|---|---|
| P | 2.34 | 1.56 | 69.1 | 568.26 | 12.05 | -45.477 |
| N | 1.05 | 0.64 | 1.1 | 23.362 | 3.508 | -190.162 |

**Table 1.** Values of different parameters used to variationally calculate the ground state energy of isolated donors within our scheme.

## 2.3. Interface Ground State $\Psi_I$

The interface Hamiltonian in presence of the single N-donor and the externally applied $F$ is written as,

$$H_I = T + V_I \quad (6)$$

$$V_I = V_\rho + V_z \quad (7)$$

$$V_\rho = \frac{2(Q-1)}{\sqrt{\rho^2+d^2}} - \frac{2}{\sqrt{\rho^2+d^2}}\left[-Ce^{-f\sqrt{\rho^2+d^2}} + g\sqrt{\rho^2+d^2}e^{-f\sqrt{\rho^2+d^2}}\right] \quad (8)$$

$$V_z = keFz - \frac{Q}{2(z+d)} \quad (9)$$

To calculate the ground state energy variationally, we minimize the Hamiltonian in Equation (6) with the normalized trial wave function,[28]

$$\Psi_I = f_\alpha(z)g_\beta(\rho) \quad (10)$$

$$f_\alpha(z) = \frac{\alpha^{5/2}}{\sqrt{4!}}(z+d)^2 e^{-\alpha(z+d)/2} \quad (11)$$

$$g_\beta(\rho) = \frac{\beta}{\sqrt{\pi}} e^{-\beta^2 \rho^2 / 2} \tag{12}$$

where, $\alpha$, $\beta$ are the variational parameters. The presence of the term $z + d$ in Equation (11) forces the ground state wave function to be terminated at the interface such that $f_\alpha(z) = 0$ for $z \leq -d$.

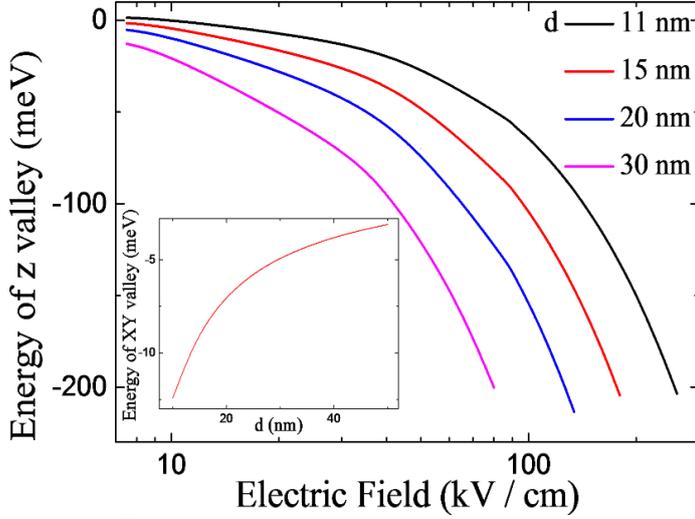

**Figure 3.** Evolution of interface $z$ valley energy levels under the influence of external electric field for four different depths of donor from the interface. The inset shows the variation of the interface $XY$ valley energy levels with $d$. The $XY$ valley is energetically very high compared to the $z$ valley asserting the electron will be always in the $z$ valley at ground state.

The ground state energy for the $z$ valley is obtained by minimizing $\langle f_\alpha(z)|T_z + V_z|f_\alpha(z)\rangle$ with respect to $\alpha$ and their modulation with external applied $F$ is shown in Figure 3 for two different donor depths from the interface. Similarly, for the $XY$ valley we minimize $\langle g_\beta(\rho)|T_\rho + V_\rho|g_\beta(\rho)\rangle$ by varying $\beta$. The variation of the energy for different depths of donor is shown in the inset of Figure 3.

Within the range of interest (100 - 200 kV/cm) for the externally applied $F$ to shuttle an electron from donor to interface, detail on that in sec. III, we can clearly observe that the $z$ valley energies are at least 100 meV lower in energy than those of $XY$ valley levels. This ensures lowest lying $z$ valley state in the interface will always serve as the electronic ground state even at room temperature ($k_B T \approx 25.8$ meV at $T \approx 300$K).

The exact energy eigen values for an infinite triangular well potential is given as,[58]

$$E_z = \left(\frac{\hbar^2}{2m_z}\right)^{1/3} \sqrt[3]{\left(\frac{3}{2}\pi eF(i + \frac{3}{4})\right)^2} \tag{13}$$

with $i = 0, 1, 2, \ldots, n$.

For an applied $F = 100$ KV/cm, the energy difference between the ground and the first excited state is $\Delta E_{01} \approx 28$ meV and for $F = 250$ kV/cm, $\Delta E_{01} \approx 53$ meV. Thus, the thermal excitation probability of the electron from interface $z$ valley ground state to the excited state are $\approx 30\%$ and $\approx 10\%$, respectively, as estimated using Boltzman distribution. Such low excitation probabilities ensure that the electron at the interface will have the maximum tendency to remain at the $z$ valley ground state, even at the room temperature. We can clearly say, higher the required $F$ to shuttle the electron from the donor to the interface, lower will be the possibility for the electron to occupy a higher $z$ valley state at the interface. This will allow us to optimize the donor position from the interface for practical charge qubit realization.

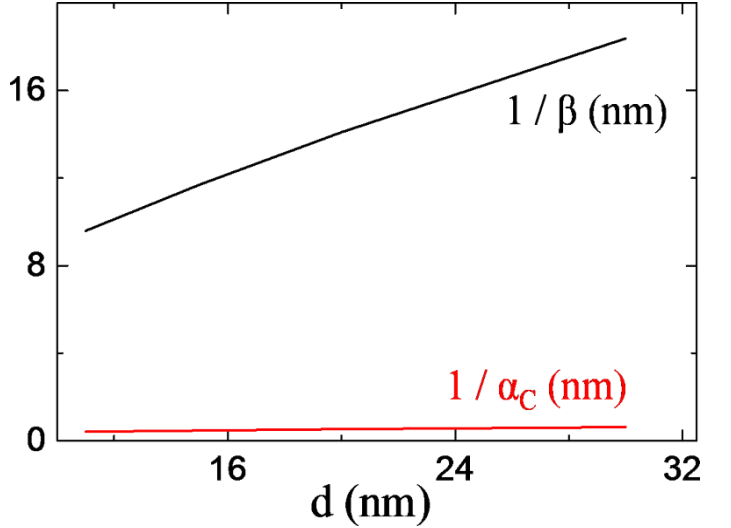

**Figure 4.** Variation of the interface $XY$ plane confinement length, $1/\beta$ (nm), and $z$ valley confinement, $1/\alpha$, against donor position. $\alpha_C$ signifies $\alpha$ for $F = F_C$ at the corresponding $d$.

The confinement lengths along the $z$ direction and in the $XY$ plane, respectively, are proportional to the inverse of the variational parameters, $1/\alpha$ and $1/\beta$. Both dependent on the depth $d$ as shown in Figure 4. $1/\alpha$ also varies with the magnitude of the applied electric field, being higher for lower $F$. Figure 4 shows the values of $1/\alpha$ at $F = F_C$ for which the donor and interface ground states are degenerate.

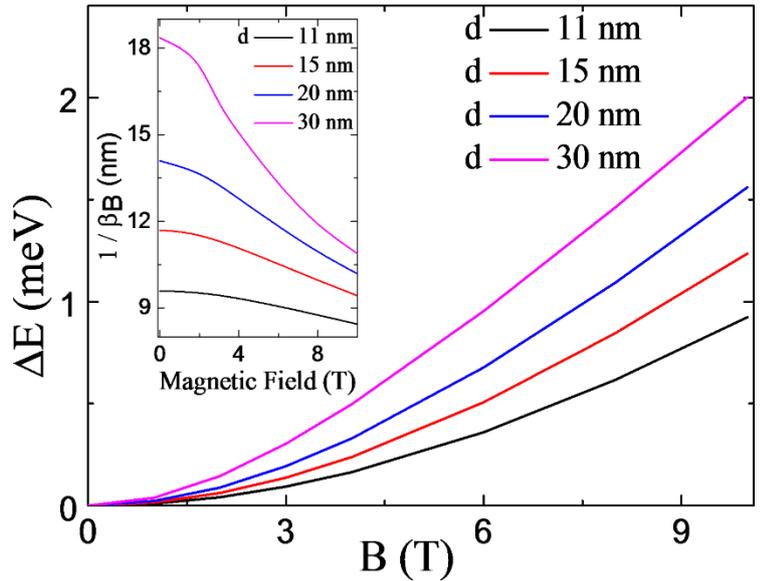

**Figure 5.** Energy shift of the interface energy levels under applied magnetic field. The inset shows variation of magnetic confinement length of the interface $XY$ levels. Higher magnetic field shrinks the electronic wave function along $XY$ plane at the interface and the kinetic energy increases subsequently.

Electronic confinement at the interface, mostly in the $XY$ plane, is highly affected by the application of an external magnetic field parallel to the $F$. Consequently, the kinetic energy increases following the presence of terms $\frac{1}{4}\mu^2\rho^2 + i\mu(y\partial_x - x\partial_y)$ in Equation (2). The extent of this phenomenon can be assessed by calculating the magnetic field, $B_C$, necessary to attain a magnetic length, $\lambda_B$, of the same order with the confinement length ($1/\beta$) in the plane parallel to the interface.

The effect of **B** related confinement is depicted in the inset of Figure 5, where variation of $1/\beta_B$ against applied $B$ is depicted for three different values of $d$. $\beta_B$ is calculated variationally by minimizing the Hamiltonian,

$$H_\rho = -\frac{\partial}{\partial x^2} - \frac{\partial}{\partial y^2} + \frac{1}{4}\mu^2\rho^2 + i\mu(y\partial_x - x\partial_y) + V_\rho \quad (14)$$

with a trial wave function $\frac{\beta_B}{\sqrt{\pi}}e^{-\beta_B^2\rho^2/2}$. The interface energy shift due to application of the external magnetic field for three different $d$'s are shown in Figure 5. It is clearly seen that the effect of external magnetic field on the interface $XY$ valley energy levels, reduces with shallower donor positioning.

## 3. Donor Interface Electron Shuttling

The application of an external **F** perpendicular to the interface (as shown in Figure 1c) develops an additional potential well, other than the donor potential, at the Si/SiO$_2$ interface. This additional triangular well energetically deepens with the increment of the external electric field. At sufficiently large values of the **F**, the energy levels at the interface become deeper in energy than the donor ground state within the system. Consequently, the donor electron ionizes and shuttles towards the interface along $z$ direction by the means of tunneling. The required field $F_C$ to shuttle the electron from donor to interface can mathematically be estimated for which $H_{II} = H_{DD}$. This is equivalent to the requirement of the two eigen-energies $E^+$ and $E^-$ of **H** to become almost degenerate. This minimum energy gap is simulated for two different depths of 11 nm and 15 nm of the single N-donor from the interface, as shown in Figure 6a and Figure 6b, respectively. The Fc values for $d = 11$ nm and $d = 15$ nm are 230 kV/cm and 161.5 kV/cm, respectively. It can be concluded that if the donor is near to the interface, $F_C$ would be higher. This is natural as the electron, even when at the interface, experiences an electrostatic pull towards the donor which restricts formation of the two-dimensional electron gas (2DEG) at the interface. The nearer the donor is to the interface, the greater will be the electrostatic pull. The variation of the $F_C$ with increasing $d$ is presented in Figure 7.

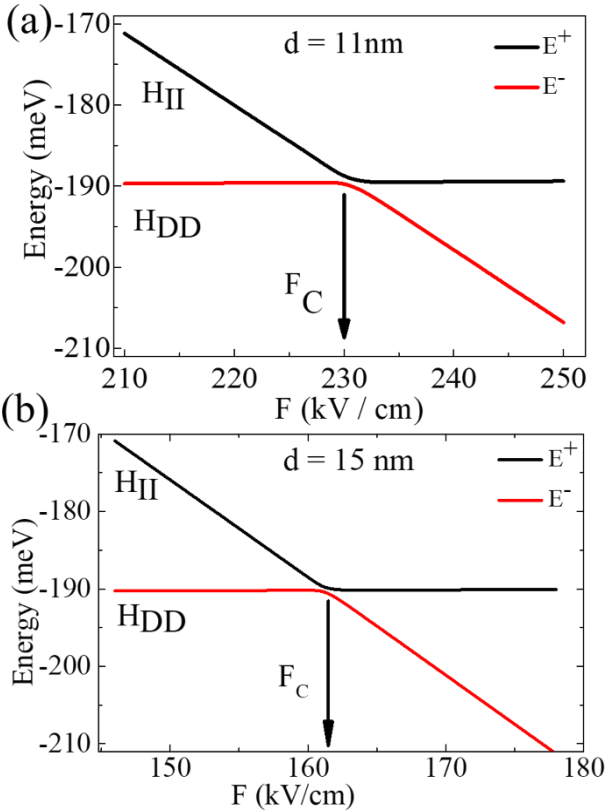

**Figure 6.** a) Evolution of the eigen-energies $E^+$ and $E^-$ with the applied electric field at $d = 11$ nm. Anti-crossing occurs at $F = F_C$ for which the minimum energy gap between $E^+$, $E^-$ is $g_{min}$. b) Anti-crossing behavior of the eigen-states at $d = 15$ nm.

For successful qubit operation at room temperature, the shuttled electron must be stable within the interface well after being shuttled from the donor atom i.e., it should not revert back to the donor due to thermal fluctuation. This in turn suggests that there should be a sufficient energy gap ($\geq k_BT = 25.8$ meV) between the two eigen-energies $E^-$ and $E^+$ beyond the degeneracy point. For example, the eigen-energies are $E^- = -189.55$ meV and $E^+ = -189.54$ meV for $d = 11$ nm at $F_C = 230$ kV/cm, as shown in Figure 6a. But, to stabilize the electron at interface against thermal fluctuation at room temperature, it could be better if $E^-$ is $\geq 220$ meV, which corresponds to an $F \geq 260$ kV/cm.

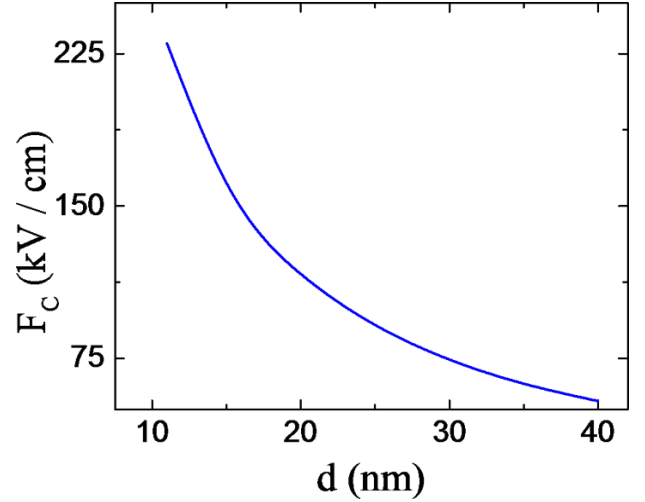

**Figure 7.** $F_c$ vs. $d$ for single N-donor placed in bulk Si.

If we consider a simple system as shown in Figure 8, we will achieve the desired uniform field $F_C$ for $d = 11$ nm, within the system by applying a gate voltage $V_0 = 8.74$V for a box layer of 100 nm using the relation,

$$V_0 = \frac{\varepsilon_1\varepsilon_2 d_3 + \varepsilon_3\varepsilon_1 d_2 + \varepsilon_3\varepsilon_2 d_1}{\varepsilon_1\varepsilon_3 d_2}(V - V_1) \quad (15)$$

Moreover, to achieve $F = 260$ kV/cm, we shall require $V_0 = 9.9$ V. For $d > 11$ nm, the external voltage required to shuttle the electron towards the interface and stabilize it there against thermal fluctuations will always be less than the voltage required for $d = 11$ nm. Such voltage values are practical to be implemented and will not damage the devices. So, it is very much reasonable to implement such a system in practice.

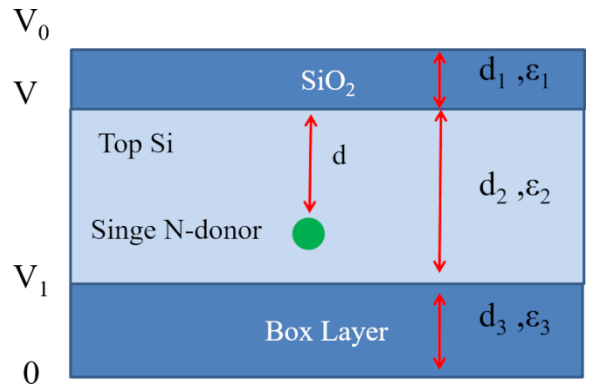

**Figure 8.** Schematic parallel plate capacitor configuration of the device to calculate for Equation (15). The bottom layer of the SOI wafer will be heavily doped silicon to make it a conducting back gate (not shown in the schematic).

The process of electron shuttling to and from the donor and the interface under the effect of externally applied $F$ and $B$ can also be demonstrated by the position expectation value of the electron. At the ground state and along $z$, the same is characterized by $<z> = \langle \psi^-|z|\psi^- \rangle$. The results for $d = 11$ nm and 15 nm are shown in Figure 9. The black horizontal lines in Figure 9a, c represent the actual position of the interface. In Figure 9a, we have simulated the evolution of the electron from the donor to the interface for $d = 11$ nm under the influence of externally applied electric field only. At small values of $F$, the electron is completely at the donor site and consequently, $\Psi^- \approx \psi_D$ and $<z> \approx 0$. Although, the center of the electron will be slightly shifted from zero due to the $(z + d)$ term that takes care of the appropriate boundary condition. Systematic increment of the applied $F$ will eventually ionize the electron from the donor and adiabatically shuttle it towards the interface. At $F = F_C$ and beyond, the electron is completely shuttled to the interface and we shall observe $\Psi^- \approx \Psi_I$ and $<z> \approx d - 5/\alpha$. This little shift in average $z$ position of the electron at the interface, after complete shuttling, is expected because $\int_{-d}^{\infty}(z+d)f_\alpha^2(z)dz = 5/\alpha$, where $\alpha$ is also field dependent. For, $F = F_C + \Delta F$, if we apply a $B$ parallel to $F$, the electron will eventually shuttle back towards the donor, anti-parallel to the electric field. This adiabatic back shuttling of electron from interface to donor at $d = 11$ nm is depicted in Figure 9b. Such behavior is obvious as the interface energy levels are affected much strongly by the external magnetic field than the donor energy levels. The synergy of parallel electric and magnetic fields offer a valuable experimental setup that aids in determining whether these charges emanate from a donor[59] or arise from some random impurities within the device, such as a metallic grain on its surface.[59,60] Similarly, the electron shuttling from the donor to interface and vice versa for $d = 15$ nm is simulated in Figure 9c, and Figure 9d, respectively. A comparison between them clearly suggests that the nearer the donor is to the interface, the electron shuttling process is more adiabatic.

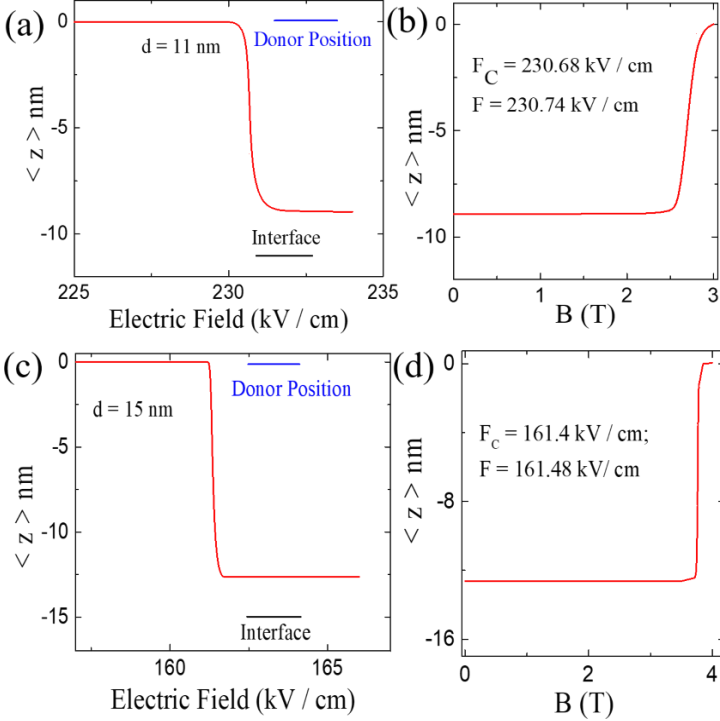

**Figure 9.** Expectation value of the electron $z$ position $<z> = \langle \psi^-|z|\psi^- \rangle$ under, a) external electric field only and b) both electric ($F = F_C + 740$ V) and magnetic field for $d = 11$ nm. c - d) for $d = 15$ nm. The nearer the donor to the interface, more adiabatic the shuttling process is.

Another very important parameter to satisfy the feasibility of charge shuttling and quantum computing in Si architecture is the electron shuttling time from the donor to the interface. It should be much smaller than the coherence time in Si. Tunneling conserves spin but compromises charge coherence. For charge qubit operation, storing quantum information, the shuttling electron must evolve adiabatically from the donor to the interface and vice-versa. Otherwise, coherence will not be conserved. Adiabatic processes involve gradual Hamiltonian changes (e.g., via external fields), sustaining the system in a known energy eigenstate throughout. This, in turn, ensures smooth progression from initial to final eigen-state.[61] Here, we mainly focus on adiabatic time over tunneling time as our main interest is charge qubit transfer at elevated temperatures, especially at room temperature regime.

The electron tunneling time is estimated as $\tau = \hbar/g_{min}$. $g_{min}$ is the minimum energy gap between the two eigen energies $E^+$ and $E^-$ (Figure 6). The adiabatic time, on the other hand, is estimated as,[62,63]

$$\tau_{ad} = \frac{\hbar |e| F_{max} d}{g_{min}^2} \quad (16)$$

$F_{max}$ is chosen to ensure the electron remains strongly in the interface, such that $\Psi^- \approx \Psi_I$. We choose $F_{max} = 2F_C$. The variation of tunneling time and adiabatic time with $d$ is shown in Figure 10.

Tunneling times are found to be ~ 1 ps for $d = 11$ nm to ~ 15 ps for $d = 20$ nm, whereas, the adiabatic times range from ~ 9.22 ns for $d = 11$ nm to ~ 225 ns for $d = 20$ nm and tend to increase rapidly for larger donor depths. For coherent shuttling of charge, these time scales are most important factors and to be compared with experimental spin and charge coherence times respectively. The spin coherence time ($T_2$) in bulk Si is of the order of few ms[6] that can further be enhanced by isotope purification[6]. Almost up to hundreds of mili-seconds. Contrary to that, presence of an surface or interface can adversely affect $T_2$.[64] In our case, even for $d = 30$ nm, $T_2/\tau \approx 10^4$ implying tunnelling time will always be shorter than spin dephasing time. But, spin qubit operation with our system at elevated temperature may be restricted due to the thermal excitation of electrons to the excited states at interface.

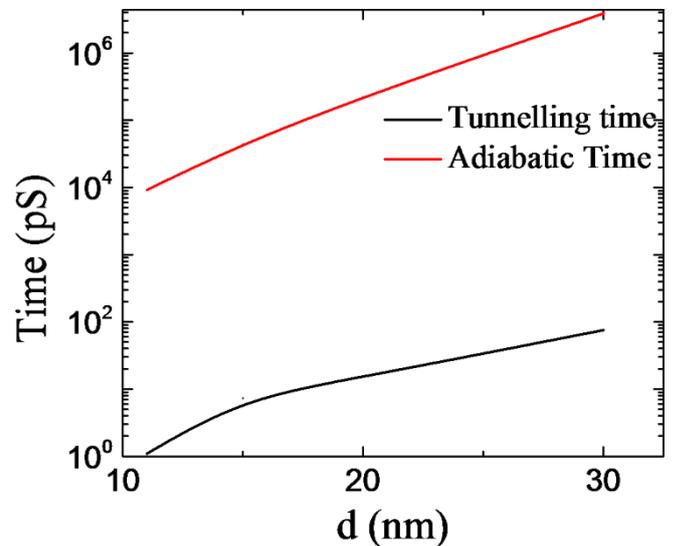

**Figure 10**: Donor electron tunneling time (black) and adiabatic passage time (red) versus donor depth $d$.

Contrary to the spin dephasing times in bulk Si, charge coherence times are much shorter. The reason for this lies in the strong coupling of electronic charges with the external environment, which is a consequence of the long-range Coulomb interaction force. The primary sources for such decoherence are either background charge fluctuations[65] within the device or electron-phonon interactions.[66] However, charge qubits are far easier to manipulate and control than spin qubits by means of external voltages. Moreover, elemental charge can also be detected precisely[18-23] even at elevated temperatures. Due to such superiority of Si charge qubit over spin qubits, ingenious spin-to-charge conversion mechanisms have also been proposed that would allow the electron-spin state to be inferred according to the absence or presence of charge detected by a SET at the device surface.[67-69] All these factors weigh heavily on the importance of a charge qubit system that would allow room temperature operation. Comparing our simulated adiabatic times (Figure10) to the experimental charge coherence time in silicon, we can say that coherent charge shuttling between donor and interface can be achieved at room temperature for such N-donor based Si devices; consequently, room temperature charge qubit transfer would be achievable.

## 4. Summary and Conclusion

Dopant atom based single electron devices have put forward an enormous opportunity to explore the fascinating realm of quantum computers. Through an extensive examination of the underlying principles, fabrication techniques, and experimental developments, the scientific community has unveiled the promising capabilities and challenges associated with this emerging technology. There have been enormous efforts put forward to rationalize practical quantum computers either by means of enhancing spin and charge coherence times or by developing robust charge qubit readout techniques at elevated temperatures. The use of hybrid systems that combine silicon-based charge qubits with other qubit technologies also adds up to the fact. Still, the bottleneck towards the generation of large-scale quantum computers remains within the lack of technology for room temperature qubit operations.

In this manuscript, we demonstrate, through extensive quantitative theoretical description, how to control and manipulate a single qubit in a doped Si architecture at room temperature. We have elaborately described how the exploitation of a single N-deep-donor within Si matrix can achieve successful room temperature electron shuttling between the donor and interface by the application of external electric and magnetic field. This, in turn, will revolutionize room temperature qubit technology. For coherent manipulation of the electron between donor and interface, the electron should always evolve adiabatically and remain on the ground state. So, we have adopted effective mass approach to have proper and precise description of the system. Further, the incorporation of the central cell potential into the Hamiltonian provides us the liberty to study such system for any arbitrary donor species. Moreover, we have discussed the stability of the charge qubit either at donor or Si/SiO$_2$ interface against the possible thermal noise at room temperature regime. We also have quantitatively described the practicality of the required external fields and device parameters, used in this study which will allow successful room temperature manipulation of charge qubit to and from the donor and interface.

Partial lifting of the Si six valley degeneracies at the interface along with easier surface electron manipulation techniques, make such devices a very good platform for qubit operation. Moreover, experiments regarding charge-qubit control in a double quantum dot at Si surface indicate that the exchange oscillatory behavior would not be a severe problem for donor-bound electrons manipulated at the Si/SiO$_2$ interface which in turn will sum up towards the possibility of multi qubit operations at room temperature with N-deep-donors within the architecture discussed.

We believe, the theoretical analysis that we carried out here, if experimentally implemented, will be a massive step forward towards room temperature quantum computing. We are on the way to experimentally realize such quantum operation.

## Acknowledgement:
Authors thank Pooja Yadav and Pooja Sudha for fruitful discussion to build the manuscript. The work is partially supported by DST-SERB (Project no: ECR/2017/001050), and IIT Roorkee (Project no: FIG-100778-PHY) India.

## Keywords
Room temperature, Quantum Bits, Deep Donor, Nitrogen Donor, Silicon FET